# REGULATING CRYPTOCURRENCY AND DECENTRALIZED FINANCE FOR AN INCLUSIVE ECONOMY

Amrutha Muralidhar
Student, Bachelor of Engineering (Computer Science and Engineering),
BMS College of Engineering, Bengaluru
amrutham.research@gmail.com *(Corresponding Author)*

Muralidhar Lakkanna
Advanced Manufacturing Research Fellow, School of Natural Sciences and Engineering,
National Institute of Advanced Studies, IISc Campus, Bengaluru

**ABSTRACT**

The evolution of cryptocurrency and decentralized finance (DeFi) marks a significant shift in the financial landscape, making it more accessible, inclusive, and participative for various societal groups. However, this transition from traditional financial institutions to DeFi demands a meticulous policy framework that strikes a balance between innovation and safeguarding consumer interests, security, and regulatory compliance. In this script we explore the imperative need for regulatory frameworks overseeing cryptocurrencies and DeFi, aiming to leverage their potential for inclusive economic advancement. It underscores the prevalent challenges within conventional financial systems, juxtaposing them with the transformative potential offered by these emergent financial paradigms. By highlighting the role of robust regulations, we examine their capacity to ensure user security, fortify market resilience, and spur innovative strides. We aim to proffer viable strategies for formulating regulatory structures that harmonize the twin objectives of fostering innovation and upholding fairness within financial ecosystems.

***Keywords:*** *Cryptocurrency, Decentralized Finance, Inclusive Economy, Traditional Finance, Regulations*

**NONMENCLATURE**

| *Abbreviations* | |
|---|---|
| AE | Advanced Economies |
| AML | Anti-Money Laundering |
| DeFi | Decentralized Finance |
| DApps | Decentralized Applications |
| EMDEs | Emerging Market and Developing Economies |
| KYC | Know-Your-Customer |
| RBI | Reserve Bank of India |
| PMLA | Prevention of Money Laundering Act |

## 1. INTRODUCTION

The financial landscape has undergone a profound shift with the rise of cryptocurrencies and Decentralized Finance (DeFi), with their market size expected to cross $39 billion by 2025[1]. These revolutionary technologies challenge traditional financial systems, offering new ways to exchange value and access financial services. This script delves into these innovations, exploring the role of regulations in steering their course and their impact on the global economy.

Cryptocurrency or Digital money along with DeFi is eliminating the need for banks as intermediaries. These platforms offer financial accessibility to a worldwide audience, reshaping how we interact with money [2]. However, alongside their promise, these innovations bring regulatory challenges. While they open doors to financial inclusion, they also pose risks that call for a delicate balance between fostering innovation and ensuring user protection. The shift from traditional finance to these decentralized systems highlights the need for robust policies that encourage progress while mitigating potential pitfalls [3]. This script examines the fundamental principles of cryptocurrencies and DeFi, investigating their divergence from conventional financial services and their growing popularity. We then study the regulatory landscape associated with these decentralized systems. By understanding the challenges faced for building a reliable regulatory framework, we aim to navigate the complex interplay between technological advancements and regulatory governance, shaping a more inclusive and resilient global financial arena.

## 2. CRYPTOCURRENCY AND DEFI

Cryptocurrencies, such as Bitcoin and Ethereum, represent a revolutionary departure from traditional fiat currency systems. These currencies operate on decentralized networks, removing the need for intermediaries like banks, and offer secure, transparent, and efficient means of transferring value [4]. DeFi utilizes blockchain and distributed ledger technology to recreate traditional financial systems in a decentralized manner [5]. DeFi platforms offer a range of financial services, such as lending, borrowing, trading, and yield farming, without relying on traditional banks or financial institutions [6]. These services are accessible to anyone with an internet connection, enabling global participation in the financial ecosystem. They have gained significant popularity in a new era of peer-to-peer financial networks, with their promise of reduced overhead costs and expedited fund transfers [7] [8]. Unlike traditional financial institutions that rely on institutional or human trust, cryptocurrencies and DeFi operate on the technology trust which has established more user trust [9]. These innovative systems have rapidly gained traction, witnessing a meteoric rise in users [9]. This surge in user adoption underscores the need for a robust policy framework that can effectively address concerns and provide a secure and trustworthy platform for all stakeholders [10]. However, the appeal of cryptocurrencies extends beyond accessibility. They serve as an alternative for individuals who face difficulties accessing traditional banking services due to financial constraints or geographical isolation. Cryptocurrencies significantly reduce the cost of financial transactions, making them more affordable for low-income individuals and small businesses. This is particularly significant in regions where traditional cross-border transactions incur exorbitant fees, acting as a major obstacle for people in developing countries. Cryptocurrencies streamline the process, enabling global transactions with minimal costs [11]. Furthermore, cryptocurrencies provide access to credit, a service often unavailable in many parts of the world where traditional banking systems are inadequate or non-existent. This financial gap is addressed through DApps built on blockchain networks. DApps like Aave and Compound Finance enable individuals to generate returns on their cryptocurrency assets or secure loans using digital holdings, all without the necessity of a conventional banking account [12] [13]. This empowers individuals to borrow money using their cryptocurrency as collateral, eliminating the constraints of traditional banking requirements. Cryptocurrencies also empower the unbanked by giving them control over their finances without the need for traditional bank accounts. Being untethered to any government or financial institution, their value hinges on supply and demand, culminating in a reliable store of value even amidst volatile markets. This feature facilitates people in protecting their savings from inflation and economic turmoil. While the decentralized nature of cryptocurrencies and DeFi seeks to eliminate the need for centralized control, it becomes evident that some degree of regulatory framework is essential to ensure user safety and platform integrity [3].

## 3. TRADITIONAL FINANCIAL INTERMEDIARIES AND INCLUSIVE ECONOMY

Traditional financial intermediaries, like banks and credit unions, facilitate the movement of capital within the financial system, serving as intermediaries between those with financial resources and those seeking them. These institutions offer various services but face challenges in bridging economic disparities, resulting in unequal economic opportunities. Problems in traditional financial systems often stem from economic frictions, such as asymmetric information, adverse selection, and moral hazard [14]. These complexities create opportunities for misuse and impose significant costs on protecting the public and the broader economy from financial fraud, misconduct, and systemic risk. Traditional financial systems come with operational costs and overheads, making financial services more expensive for end-users, especially those in underserved or disadvantaged segments of the population [15]. This exacerbates economic disparities, restricts access to financial resources, and discourages participation in formal financial systems, hindering the goal of economic inclusion.

As these problems persist, Cryptocurrencies and DeFi have emerged as promising alternatives that

 

promote economic inclusion. Their use of smart contracts enhances efficiency by automating transactions, ensuring accuracy with code-based execution, and providing immutability for tamper-proof agreements [16]. They reduce dependency on third parties, enabling secure and transparent transactions without intermediaries. Although, smart contracts are permanent and demand precision in programming [17]. They bridge economic disparities, providing accessible, equitable, and efficient financial services, promoting global economic stability and inclusion. Also, cryptocurrencies are programmable currencies that scale remarkably by using payment channels like the Lightning Network. This significantly increases transaction throughput over traditional centralized networks [18]. Figure 1 gives the pictorial understanding of decentralized financial networks and traditional financial networks.

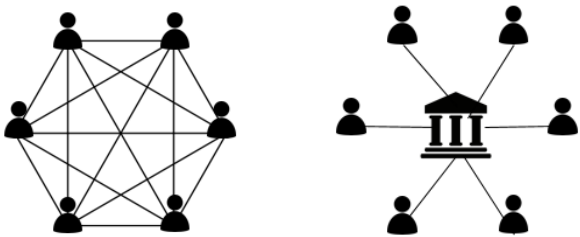

*Figure 1:* Decentralized finance network and Traditional finance network

In this script we consider economic inclusion as providing fundamental financial services to all segments of society. It aims to empower historically excluded or underserved individuals and communities, reducing disparities and promoting economic stability. By granting access to financial services, economic inclusion helps individuals and businesses secure their financial futures, break free from poverty, and contribute to economic growth [19].

## 4. REGULATORY FRAMEWORK IN INDIA

In India KYC, and Prevention of Money Laundering Act, 2002 [20] regulations are applied to cryptocurrencies and DeFi to mitigate financial crimes. These regulations require individuals and entities involved in cryptocurrency transactions and DeFi activities to provide their identity. Exchanges and platforms dealing with cryptocurrencies and DeFi applications in India are typically obligated to implement KYC procedures with verifiable user identities. PMLA laws mandate the reporting of suspicious transactions and the implementation of monitoring mechanisms to detect and prevent money laundering activities within the cryptocurrency sector [21].

The Reserve Bank of India has adopted a cautious approach to cryptocurrencies in recent years. In 2018, RBI issued a circular that effectively prohibited regulated financial institutions from providing services to crypto-related businesses and individuals [22]. However, this circular was later overturned by the Supreme Court of India in 2020, allowing banks to re-engage the cryptocurrency sector [23]. In June 2019, a government-appointed committee proposed a draft bill titled the "Banning of Cryptocurrency and Regulation of Official Digital Currency Bill, 2019." This bill recommended a complete ban on cryptocurrencies in India, with provisions for the development of a central bank digital currency (CBDC) [24]. The government has since displayed a more nuanced stance on cryptocurrencies, indicating the intention to explore the potential benefits of blockchain technology while continuing to address the challenges posed by digital currencies. In a report published on June 2023, RBI lays forth three possible policy responses to address the risks associated with cryptocurrencies: ban, containment, or regulation. It emphasizes the need for a globally coordinated approach to analyze risks posed to EMDEs compared to AEs, especially concerning macroeconomic challenges, development challenges, and cross-border challenges. The report also mentions that there is a need to create a global regulatory framework for unbacked crypto assets, stablecoins, and DeFi while considering both macroeconomic and regulatory perspectives [25].

## 5. REGULATORY CHALLENGES

Cryptocurrencies and DeFi applications, built on permissionless protocols, prioritize user anonymity and typically do not collect personal information about their users [26]. While this emphasis on privacy can be seen as an advantage, it also presents a range of challenges:

1. Enforcement of KYC/AML Laws: Cryptocurrency transactions are often pseudonymous, meaning that they do not directly reveal the identities of the participants. While blockchain records all transactions, it typically only shows the sender and receiver's wallet addresses, making it challenging for regulatory authorities to verify user identities [27]. As they operate across borders and transact globally, it becomes challenging for local regulatory authorities to enforce

 

consistent regulatory protocols. The lack of universal KYC / AML protocol for the cryptocurrency transaction across exchanges and businesses is causing multiple issues from a compliance perspective.

2. Tax Compliance: Cryptocurrency transactions can be challenging to trace, making it difficult for tax authorities to ensure that individuals and businesses accurately report their cryptocurrency holdings and transactions for tax purposes [28]. This lack of transparency can result in potential tax revenue losses.

3. Financial Crime and Fraud: The anonymity provided by blockchain technology can facilitate various financial crimes, like fraud, money laundering, and terrorist financing [29]. Hence, there is a challenge of developing effective mechanisms for detecting and preventing these illicit activities.

4. Consumer Protection: Cryptocurrency transactions are often irreversible, which can leave consumers vulnerable to scams, hacks, and the loss of their digital assets [30]. There is a need to find innovative ways to protect consumers in this relatively new and rapidly evolving financial landscape.

5. International Coordination: Cryptocurrencies inherently operate across borders [31], and the lack of international consensus on regulatory approaches makes it challenging to address these challenges effectively. Harmonizing regulations and standards across different jurisdictions is a complex and ongoing task.

6. Rapid Technological Evolution: Cryptocurrencies and blockchain technology are continually evolving, introducing new features and capabilities. Forming regulations that keep up with these innovations and adapt their regulatory approaches accordingly to maintain effective oversight is also a challenging task.

## 6. IMPACT

Well-structured regulations play a multifaceted role in the cryptocurrency and DeFi landscape, enhancing various aspects of the market. These regulations provide much-needed legitimacy, fostering investor confidence and safeguarding their interests. They serve as a catalyst for economic inclusion and stimulate market growth by attracting investments, spurring innovation, and expanding the cryptocurrency and DeFi ecosystem. Regulatory harmonization promotes seamless cross-border transactions, creating a conducive environment for market participants [32]. It's instrumental in combating financial crimes and mitigating the risks associated with the relative anonymity of cryptocurrencies. This also attracts institutional investors who prioritize clarity and protection when entering new markets. This institutional adoption contributes to market stability and the inflow of capital. For example, the Securities and Exchange Commission has provided guidance on the regulatory framework for initial coin offerings and token sales. This guidance has helped to create a more stable environment for investors and has contributed to the growth of the industry [33]. Forward-thinking regulations serve as a driving force behind technology advancement in blockchain and DeFi, facilitating the creation of improved financial products and services accessible to a wider demographic. Transparent and enforceable regulations instill consumer confidence by providing security and trust in cryptocurrency and DeFi platforms. Striking the right balance between regulation and innovation is critical, ensuring investor protection, curbing fraud, stimulating innovation, and propelling the growth and adoption of these transformative technologies.

## 7. CONCLUSION

Our study highlights how cryptocurrencies and DeFi are reshaping the financial landscape and promoting greater inclusion and accessibility. We've outlined the challenges within traditional finance and the pivotal role regulations play in ensuring safety and fostering innovation in these dynamic systems. Effective regulation, rooted in an understanding of these technologies, societal needs, and economic objectives, holds the key to an inclusive, resilient, and fair financial future. Achieving an inclusive economy through regulated cryptocurrency and DeFi systems necessitates a delicate equilibrium. It calls for embracing innovation while ensuring accountability, protection, and stability for all stakeholders.